\documentclass[aps, prd, showpacs, superscriptaddress, ctexart, nofootinbib, twocolumn]{revtex4}

\usepackage{amssymb, amsmath, bm, dcolumn, epsf, graphicx, latexsym, slashed, simplewick}

\usepackage{color}
\newcommand{\tc}{\textcolor{black}}
\def\be{\begin{equation}}
\def\ee{\end{equation}}
\def\bea{\begin{eqnarray}}
\def\eea{\end{eqnarray}}
\begin{document}

\title{Emergent Universe Scenario via Quintom Matter}

\author{Yi-Fu Cai\footnote{Email: ycai21@asu.edu}}
\affiliation{Department of Physics, McGill University, Montr\'eal, QC, H3A 2T8, Canada}
\affiliation{Department of Physics, Arizona State University, Tempe, AZ 85287, USA}

\author{Mingzhe Li\footnote{Email: limz@nju.edu.cn}}
\affiliation{Department of Physics, Nanjing University, Nanjing 210093, China}
\affiliation{Joint Center for Particle, Nuclear
Physics and Cosmology, Nanjing University - Purple Mountain
Observatory, Nanjing 210093, China}

\author{Xinmin Zhang\footnote{Email: xmzhang@ihep.ac.cn}}
\affiliation{Institute of High Energy Physics, Chinese Academy of Sciences, Beijing 100049, China}


\begin{abstract}
The emergent universe scenario provides a possible alternative to bouncing cosmology to avoid the Big Bang singularity problem. In this paper we study the realization of the emergent universe scenario by making use of Quintom matter with an equation of state across the cosmological constant boundary. We will show explicitly the analytic and numerical solutions of emergent universe in two Quintom models, which are a phenomenological fluid and a nonconventional spinor field, respectively.
\end{abstract}

\maketitle

\section{Introduction}

Inflation is considered as the most successful model of describing physics of very early universe, which has explained conceptual issues of the Big Bang cosmology \cite{Guth:1980zm, Linde:1981mu, Albrecht:1982wi} (see \cite{Starobinsky:1980te, Fang:1980wi, Sato:1980yn} for early works). Among these remarkable achievements, inflation has predicted a nearly scale-invariant primordial power spectrum which was later verified in high precision by Cosmic Microwave Background (CMB) observations \cite{Komatsu:2010fb}. The success of inflation is mainly based on a series of assumptions including an enough long period of quasi-exponential expansion and the applicability of perturbation theory during this phase. However, these assumptions often bring troubles to inflation models, such as the fine-tuning problem of the potential parameters. Moreover, it was pointed out that inflation models suffer from the initial singularity problem inherited from the Big Bang cosmology \cite{Borde:1993xh}.

Recently, there are increasing interests in alternatives to inflationary cosmology which can not only be the same successful as inflation in explaining early universe physics but also avoid the initial spacetime singularity (for example see \cite{Brandenberger:2011gk} for a recent review). One class of the alternative scenario is the bouncing cosmology, which suggests the expansion of our universe was preceded by an initial contraction and then a non-vanishing bouncing point happened to connect the contraction and the expansion \cite{Mukhanov:1991zn, Brandenberger:1993ef, Cai:2012va}. It was pointed out in Ref. \cite{Cai:2007qw}
that a realistic cosmological model realizing a nonsingular bounce requires a matter field with Quintom behavior \cite{Feng:2004ad} of which the equation of state (EoS) of the universe has to cross over the cosmological constant boundary $w=-1$ twice. This type of bounce model was extensively studied in the literature in recent years (for example see \cite{Cai:2007zv, Cai:2008ed, Cai:2008qb, Cai:2009hc} and references therein, and see \cite{Cai:2009zp} for a review on Quintom cosmology). A bouncing model combining the {\it Matter Bounce} scenario \cite{Wands:1998yp, Finelli:2001sr} and Quintom scenario \cite{Feng:2004ad} was achieved in the frame of Lee-Wick cosmology \cite{Cai:2008qw}, which showed that such kind of bounces can give rise to a scale-invariant power spectrum for primordial curvature perturbation. Later, the cosmological perturbation theory of bouncing cosmology was established to a complete frame which includes primordial non-gaussianities \cite{Cai:2009fn, Cai:2009rd}, entropy fluctuations and curvaton scenario \cite{Cai:2011zx}, and related preheating process \cite{Cai:2011ci}. A recent nonsingular bouncing model \cite{Cai:2012va} which combined the benefit of {\it Matter Bounce} \cite{Wands:1998yp, Finelli:2001sr} and {\it Ekpyrotic} cosmology \cite{Khoury:2001wf, Lehners:2008vx} was proposed and the $w$ crossing $-1$ scenario can be realized without pathologies through the Galileon-like Lagrangian (see \cite{Nicolis:2008in} for the Galileon model and \cite{Deffayet:2011gz, Horndeski:1974} for extensions).

Besides the bouncing, there is another interesting cosmological scenario which could be a strong competitor of inflation, which is the so-called emergent universe scenario \cite{Brandenberger:1988aj, Ellis:2002we, Ellis:2003qz}. This scenario suggests that our universe was initially emergent from a non-vanishing minimal radius and experienced a enough long quasi-Minkowski phase and then entered the normal thermal expansion. It was obtained in the {\it String Gas} cosmology in which the emergent universe was achieved in the Hagedorn phase of a thermal system composed of a gas of superstrings \cite{Brandenberger:1988aj, Battefeld:2005av, Brandenberger:2008nx}. It can be implemented by a model dubbed as {\it Galilean Genesis} as well \cite{Creminelli:2010ba}. Phenomenological study on the causal generation of primordial field fluctuations in this scenario was performed via the so-call {\it conformal cosmology} \cite{Rubakov:2009np, Osipov:2010ee, Libanov:2010nk, Libanov:2010ci, Libanov:2011bk, Libanov:2011zy}, and the {\it pseudo-conformal cosmology} \cite{Hinterbichler:2011qk, Hinterbichler:2012mv}. Later, the issue of successfully transferring scale-invariant primordial field fluctuations to curvature perturbations was discussed in \cite{LevasseurPerreault:2011mw} and \cite{Wang:2012bq}, respectively. Moreover, there is a modified version of emergent universe in which the universe has experienced a process of slow contraction \cite{Khoury:2001wf, Khoury:2009my, Khoury:2010gw, Joyce:2011ta} or slow expansion \cite{Piao:2003ty, Piao:2004tq, Piao:2007sv, Piao:2010bi, Liu:2011ns, Liu:2012ww}. {\tc{Recently, it has been also shown that emergent-type universes, which start out with a very small Hubble rate, are quite generally preferred over inflationary models in a landscape, as described in \cite{Lehners:2012wz, Johnson:2011aa}.}}

In this paper, we are going to show that within the frame of the 4-dimensional Friedmann-Robertson-Walker (FRW) universe governed by standard Einstein gravity, the Quintom-like matter field is needed to realize a realistic model of emergent universe, which drives the universe from a quasi-Minkowski phase to a normal thermal expanding history. We start with a brief examination on this necessary condition.

Consider a universe initially emergences from a nonzero minimal size and experiences enough long period of quasi-Minkowski phase. In this phase, the scale factor $a(t)$ is almost a constant and its time derivative $\dot{a}(t)$ is nearly zero. In order to make the universe exit this phase gradually, we need its second order time derivative $\ddot{a}(t)$ to be a very small positive value. Thus, if we use the Hubble parameter $H$
($\equiv\dot{a}/a$) to characterize dynamics of this phase, we find $H \rightarrow 0^+$ and $\dot{H} > 0$. Having assumed Einstein gravity is still available to describe the gravity sector during this period, one can immediately read there is an effective EoS of the universe which satisfies,
\begin{eqnarray}
 w = -1-\frac{2\dot{H}}{3H^2} \ll -1~.
\end{eqnarray}
After exiting the quasi-Minkowski evolution gracefully, the universe needs to enter into the normal thermal expanding phase which requires the EoS of the whole universe to be roughly equal to $1/3$, $0$ and $-1$ along with the observable history, respectively. Therefore, this requires a transit of the background EoS from $w<-1$ to $w>-1$, which is exactly the Quintom behavior.

In this paper we study particular realizations of the emergent universe picture by using several explicit Quintom models. The paper is organized as follows. In Section II, we study the general requirements of the emergent universe picture. Then in Section III, we present the analytic and numerical solution of the emergent universe from a toy model of a phenomenological Quintom-like fluid with a parameterized EoS across the cosmological constant boundary. For a much concrete model building, we consider a nonconventional spinor field \cite{Cai:2008gk} in Section III. Particularly we make use of an ansatz of the EoS and reconstruct the potential of the spinor Quintom analytically and numerically. Section V is devoted to a summary of the paper.

\section{The requirements of the emergent universe}

To start, we consider a spatially flat FRW universe, of which the metric is given by
\begin{eqnarray}\label{element}
 ds^2 = dt^2 -a^2(t) d\vec{x} \cdot d\vec{x}~.
\end{eqnarray}
Without modifying General Relativity, it is well known that the background dynamics in this frame follow the following two equations of motion,
\begin{eqnarray}
 H^2 = \frac{\rho}{3M_p^2}~,~~\dot{H} = -\frac{\rho+P}{2M_p^2}~,
\end{eqnarray}
in which $M_p\equiv 1/\sqrt{8\pi G}$ is the reduced Planck mass. In addition, $\rho$ and $P$ are the energy density and the pressure of the matter fields filled in the universe, respectively.

The emergent universe is a scenario of non-singular cosmology. Differing from the bouncing or cyclic modes where the expanding universe was preceded by a contracting phase, the picture of emergent universe requires the universe expands forever beginning with a finite scale factor in the infinite past. This implies the Hubble rate $H$ cannot be negative.

To be non-singular requires the spacetime of the universe is geodesically complete, i.e., the affine parameters of geodesics are divergent in the limit of infinite past \cite{Borde:2001nh}. The null geodesic obeys the equations
\begin{eqnarray}
 \frac{dk^{\nu}}{d\lambda}+\Gamma^{\nu}_{\mu\rho}k^{\mu}k^{\rho}=0~,
\end{eqnarray}
and
\begin{eqnarray}
 g_{\mu\nu}k^{\mu}k^{\nu}=0~,
\end{eqnarray}
where $\lambda$ is the affine parameter and $k^{\mu}=dx^{\mu}/d\lambda$ is the vector tangential to the geodesic. In the spatially flat FRW universe with the line element (\ref{element}), one can show that
\begin{eqnarray}
 \frac{dk^0}{dt}+Hk^0=0~.
\end{eqnarray}
By making use of $H=d\ln a/dt$, the above equation yields
\begin{eqnarray}
 k^0=\frac{dt}{d\lambda}\propto {1\over a}~,~d\lambda \propto a(t) dt
\end{eqnarray}
and hence, the requirement of the completion of null geodesics can be expressed as
\begin{eqnarray}
 |\int_{t'=-\infty}^{t'=t} a(t')dt'|=|\int_{a(t'=-\infty)}^{a(t'=t)}H^{-1}da| = \infty~.
\end{eqnarray}
Because the scale factor $a$ is finite and non-negative, the divergence of the second integral needs $H^{-1}$ should be singular at certain moment. For example, in bouncing cosmology, $H^{-1}=\infty$ at the bouncing point which is located at some intermediate time. However, for the emergent universe scenario, it is natural to conclude that $H^{-1} \rightarrow \infty$ when $t'\rightarrow -\infty$, which means, the spacetime of the universe approaches Minkowskian in the infinite past.

For time-like geodesics, we can think that they are worldlines of some free particles with masses. Consider a wordline of a free particle with mass $m$, the proper time $s$ itself is the affine parameter of the wordline. The four-momentum is defined as
\begin{eqnarray}
 P^{\mu}=m\frac{dx^{\mu}}{ds}~,
\end{eqnarray}
which obeys $g_{\mu\nu}P^{\mu}P^{\nu}=m^2$. The condition for the geodesic completion is obtained from the geodesic equation, which is
\begin{eqnarray}
 |\int_{a(t'=-\infty)}^{a(t'=t)}H^{-1}(a^2+\frac{C_p}{m^2})^{-1/2}da|=\infty~,
\end{eqnarray}
where $C_p=P_iP_i$ is a non-negative constant. This also requires the universe starts at infinite past from a Minkowski spacetime where $H^{-1}=\infty$ or $H=0$. The evolution of a Minkowski spacetime to an expanding universe requires the $H$ increases for some time during which $\dot H>0$, this implies the equation of state $w<-1$ and the null energy condition was violated. At later time the universe should enter into the radiation dominated phase in which $w=1/3$. So $w$ crosses $-1$ at some intermediate time. That is to say, the matter dominating the universe has Quintom behavior.

Furthermore, in the emergent universe the Hubble rate reached a maximum $H_{max}$ at the crossing point. It defines a mass scale the universe cannot go beyond. If $H_{max}$ is much smaller than the Planck mass, the validity of the general relativity as a low energy effective theory is guaranteed and the corrections of quantum gravity are suppressed.

\section{Emergent Universe with a Quintom-like fluid}

As a first example, we illustrate the possibility of obtaining the emergent universe solution in a phenomenological Quintom fluid described by the following EoS:
\begin{eqnarray}\label{EoS1}
 w(t) = \frac{1}{3}-\frac{2\alpha}{3}e^{-\alpha M_p t}~,
\end{eqnarray}
where $\alpha$ is a positive-valued dimensionless parameter. In this particular parametrization, we have assumed the universe can automatically enters the radiation dominated period after the quasi Minkowski expansion. Thus, one can see the EoS approaches $1/3$ when the cosmic time $t$ goes to positive infinity. Moreover, when $t$ approaches to a far past moment, the EoS would have fallen down to negative infinity very soon due to the expression of the exponential term.

\begin{figure}
\includegraphics[scale=0.4]{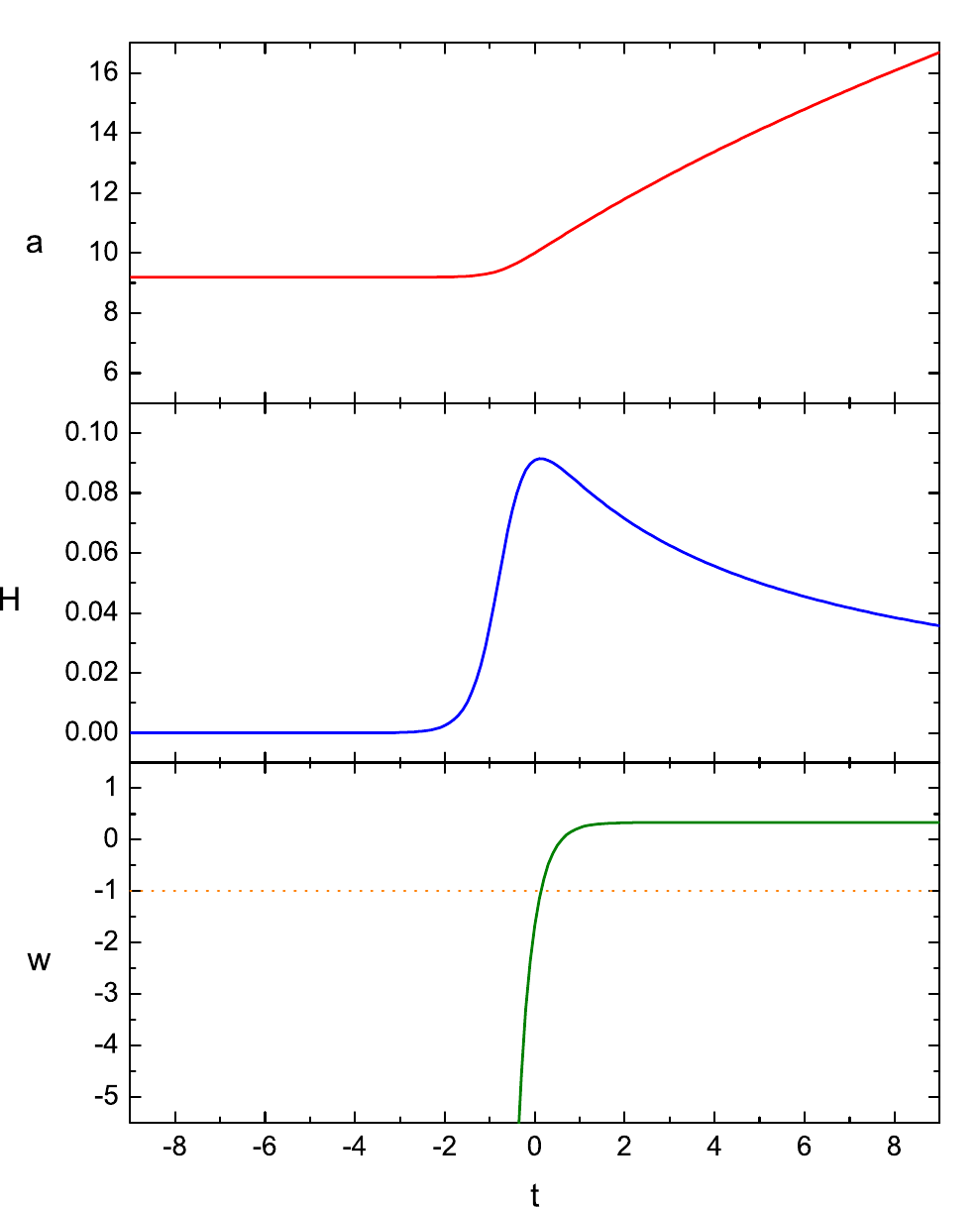}
\caption{Plot of the evolutions of the scale factor $a$, the Hubble parameter $H$ and the background EoS $w$ as a function of cosmic time in the case of Quintom fluid. In the numerical calculation, we take $\alpha=3$ and $C=10$. All dimensional parameters are of Planck units. }
\label{Fig:emergent1}
\end{figure}

To substitute the parametrization of the EoS (\ref{EoS1}) into the Friedmann equations, one can solve out the explicit solution to the Hubble parameter as follows,
\begin{eqnarray}
 H(t) = \frac{M_p e^{\alpha M_p t}}{1+ (C+2M_pt) e^{\alpha M_p t}}~,
\end{eqnarray}
where $C$ is a constant which can determine the energy scale of the Hubble parameter at the occurrence of emergent universe. Further, one obtains the energy density and the pressure of the Quintom fluid governing the universe as well. As a consequence, the evolution of the scale factor in this model can be numerically integrated out as shown in Fig. \ref{Fig:emergent1}. From the figure, one can explicitly read that the scale factor $a$ approaches to a non-vanishing minimal value in the limit of far past, and connects the thermal expansion when $t\rightarrow +\infty$.

Notice that, the occurrence moment $t_E$ of the emergent universe scenario can be characterized by the moment of $w$ crossing $-1$. Thus, by solving $w(t_E)=-1$, one gets
\begin{eqnarray}
 t_E = \frac{1}{\alpha M_p} \log(\frac{\alpha}{2}) ~,
\end{eqnarray}
which in our explicit example takes the value of $0.14$ approximately. At this moment, the Hubble parameter arrives at the maximal value
\begin{eqnarray}
 H_E = \frac{M_p}{C+\frac{2}{\alpha}(1-\log(\frac{2}{\alpha}))}~,
\end{eqnarray}
which is around $0.1 M_p$. Therefore, we are able to trust Einstein gravity in this case.

\section{Emergent Universe and spinor Quintom}

In this section, we consider a class of Quintom model described by a nonconventional spinor field. To begin with, we simply review the background dynamics of a spinor field which is minimally coupled to Einstein's gravity (see Refs. \cite{Weinberg, BirrellDavies} for detailed introduction and see \cite{ArmendarizPicon:2003qk, Cai:2008gk, Feng:2012jm} for recent phenomenological study in cosmology).

\subsection{Algebra of a cosmological spinor}

Following the general covariance principle, a connection between the metric $g_{\mu\nu}$ and the vierbein is given by
\begin{equation}
 g_{\mu\nu}e_{a}^{\mu}e_{b}^{\nu}=\eta_{ab}~,
\end{equation}
where $e_{a}^{\mu}$ denotes the vierbein, $g_{\mu\nu}$ is the space-time metric, and $\eta_{a b}$ is the Minkowski metric with $\eta_{ab}={\rm diag}(+1,-1,-1,-1)$. Note that the Latin indices represents the local inertial frame and the Greek indices represents the space-time frame.

We choose the Dirac-Pauli representation as
\begin{eqnarray}
 \gamma^0= \left(\begin{array}{cccc}
 1 &   0  \\
 0 &   -1
 \end{array}\right),~~~
 \gamma^{i}=\left(\begin{array}{cccc}
 0 &          \sigma_{i} \\
 -\sigma_{i}&   0
 \end{array}\right),
\end{eqnarray}
where $\sigma_{i}$ is Pauli matrices. One can see that the $4\times4$ $\gamma^{a}$ satisfy the Clifford algebra $\{\gamma^{a},\gamma^{b}\}=2\eta^{ab}$. The $\gamma^{a}$ and $e_{a}^{\mu}$ provide the definition of a new set of Gamma matrices
\begin{equation}
 \Gamma^{\mu}=e_{a}^{\mu}\gamma^{a}~,
\end{equation}
which satisfy the algebra $\{\Gamma^{\mu},\Gamma^{\nu}\}=2g_{\mu\nu}$. The generators of the Spinor representation of the Lorentz group can be written as $\Sigma^{ab}=\frac{1}{4}[\gamma^{a},\gamma^{b}]$. So the covariant derivative are given by
\begin{eqnarray}
 D_{\mu}\psi&=&(\partial_{\mu}+\Omega_{\mu})\psi~,\\
 D_{\mu}\bar\psi&=&\partial_{\mu}\bar\psi-\bar\psi\Omega_{\mu}~,
\end{eqnarray}
where the Dirac adjoint is defined as $\bar\psi\equiv\psi^+\gamma^0$. The $4\times4$ matrix $\Omega_{\mu} = \frac{1}{2}\omega_{\mu ab}\Sigma^{ab}$ is the spin connection, where $\omega_{\mu ab} = e_{a}^{\nu} \nabla_{\mu}e_{\nu b}$ are Ricci spin coefficients.

By the aid of the above algebra we can write down the following Dirac action in a curved space-time background
\begin{eqnarray}\label{action}
 S_{\psi} = \int d^4 x~e~ [\frac{i}{2}(\bar\psi\Gamma^{\mu}D_{\mu} \psi-D_{\mu}\bar\psi\Gamma^{\mu}\psi)-V(\bar\psi\psi)]~.
\end{eqnarray}
Here, $e$ is the determinant of the vierbein $e_{\mu}^{a}$ and $V$ stands for the potential of the spinor field $\psi$ and its adjoint $\bar\psi$. Due to the requirement of covariance, the potential $V$ only depends on the scalar bilinear $\bar\psi\psi$ and ``pseudo-scalar" term $\bar\psi\gamma^5\psi$. For simplicity we drop the latter term and only assume $V=V(\bar\psi\psi)$.

Varying the action with respect to the vierbein $e_{a}^{\mu}$, we obtain the energy-momentum-tensor,
\begin{eqnarray}\label{EMT}
 T_{\mu\nu}&=&\frac{e_{\mu a}}{e}\frac{\delta S_\psi}{\delta e_{a}^{\nu}} \nonumber\\
 &=& \frac{i}{4} [\bar\psi\Gamma_{\nu}D_{\mu}\psi +\bar\psi \Gamma_{\mu}D_{\nu}\psi -D_{\mu}\bar\psi\Gamma_{\nu}\psi -D_{\nu}\bar\psi\Gamma_{\mu}\psi]
 \nonumber\\
 && -g_{\mu\nu}{\cal L}_{\psi}~.
\end{eqnarray}
On the other hand, varying the action with respect to the field $\bar\psi$, $\psi$ respectively yields the following equations of motion,
\begin{eqnarray}
\label{EoMpsi}
 i\Gamma^{\mu}D_{\mu}\psi-V'\psi &=& 0~,\\
\label{EoMbarpsi}
 iD_{\mu}\bar\psi\Gamma^{\mu}+V'\bar\psi &=& 0~,
\end{eqnarray}
where $V'\equiv\partial V/\partial({\bar\psi\psi})$ denotes the derivative of the spinor potential with respect to $\bar\psi\psi$.

We deal with the homogeneous and isotropic FRW metric. Correspondingly, the vierbein are given by
\begin{equation}
 e_{0}^{\mu}=\delta_{0}^{\mu}~,~~e_{i}^{\mu}=\frac{1}{a}\delta_{i}^{\mu}~.
\end{equation}
Assuming the spinor field is space-independent, the equation of motion reads $ i\gamma^{0} (\dot{\psi} + \frac{3}{2}H\psi) - V' \psi = 0 $, where a dot denotes a derivative with respect to the cosmic time and $H$ is the Hubble parameter. Taking a further derivative, we can obtain:
\begin{equation}\label{solution}
 \bar\psi\psi=\frac{N}{a^{3}}~,
\end{equation}
where $N$ is a positive time-independent constant.

From the expression of the energy-momentum tensor in Eq. (\ref{EMT}), one can obtain the expressions of the energy density and the pressure of the spinor field as follows,
\begin{eqnarray}
\label{density}\rho_{\psi}&=&T_{0}^{0}=V~,\\
\label{pressure}p_{\psi}&=&-T_{i}^{i}=V'\bar\psi\psi-V~.
\end{eqnarray}
As a consequence, the EoS of the spinor field is given by
\begin{equation}\label{eos}
 w_{\psi}\equiv\frac{p_{\psi}}{\rho_{\psi}}=-1+\frac{V'\bar\psi\psi}{V}~.
\end{equation}

\subsection{Reconstruction of Spinor Quintom realizing emergent universe scenario}

The above formulae show that a cosmological spinor field might realize its EoS to cross the cosmological constant boundary when the sign of $V'$ changes. For a conventional spinor with its potential taking the form of $m\bar\psi\psi$, its EoS is exactly zero which coincides with that of normal non-relativistic dust matter. Thus, one has to consider a nonconventional form of the potential for the cosmological spinor to achieve
the Quintom scenario.

Interestingly, observing the form of the Friedmann equations and the formulae of the spinor field, we can see that a model of spinor Quintom can be reconstructed to fulfill the scenario of emergent universe. Suppose a fixed form of the EoS such as what we have illustrated in the case of Quintom fluid. We then read the expression of the Hubble parameter $H$ and its time derivative $\dot{H}$, and thus can further derive the evolutions of $V$ and $V'$ as functions of cosmic time. By making use of the relation that $\bar\psi\psi \propto a^{-3}$, the form of $V$ can be derived out explicitly. We will do the reconstruction in the following context.

Note that, a realistic cosmological evolution requires a thermal expansion after the primordial period. After radiation dominated phase, the universe enters into a matter dominated era. However, we usually do not expect that a fermion field could be responsible for a large amount of radiation \footnote{During radiation dominated phase, the main contribution comes from the gauge photons, and only a few part is from nearly massless hot neutrinos.}. Therefore, for a simple and natural choice, we parameterize the form of EoS for the spinor field as follows,
\begin{eqnarray}\label{EoSpsi}
 w_{\psi} = -\frac{2\alpha}{3} e^{-\alpha M_p t} ~.
\end{eqnarray}
To observe this form, one can see that the EoS falls into negative infinity when $t\ll-1$, but approaches $0$ at $t\rightarrow+\infty$. This parametrization can give rise to a emergent universe solution with a dust-like expansion following the quasi-Minkowski phase without the radiation domination.

Substituting Eq. (\ref{EoSpsi}) into the Friedmann equations, one can solve out the Hubble parameter as a function of cosmic time:
\begin{eqnarray}\label{H_Emergent}
 H(t) = \frac{M_p e^{\alpha M_p t}}{1+ (C+\frac{3}{2}M_pt) e^{\alpha M_p t}}~,
\end{eqnarray}
and thus the potential of the spinor evolves as
\begin{eqnarray}
 V = \frac{12M_p^4 e^{2\alpha M_p t}}{(2+ 2(C+3M_pt) e^{\alpha M_p t})^2}~.
\end{eqnarray}
Similar to the case of Quintom fluid, the coefficient $C$ is a integral constant which is used to determine the energy scale of the occurrence of emergent universe scenario.

\begin{figure}
\includegraphics[scale=0.4]{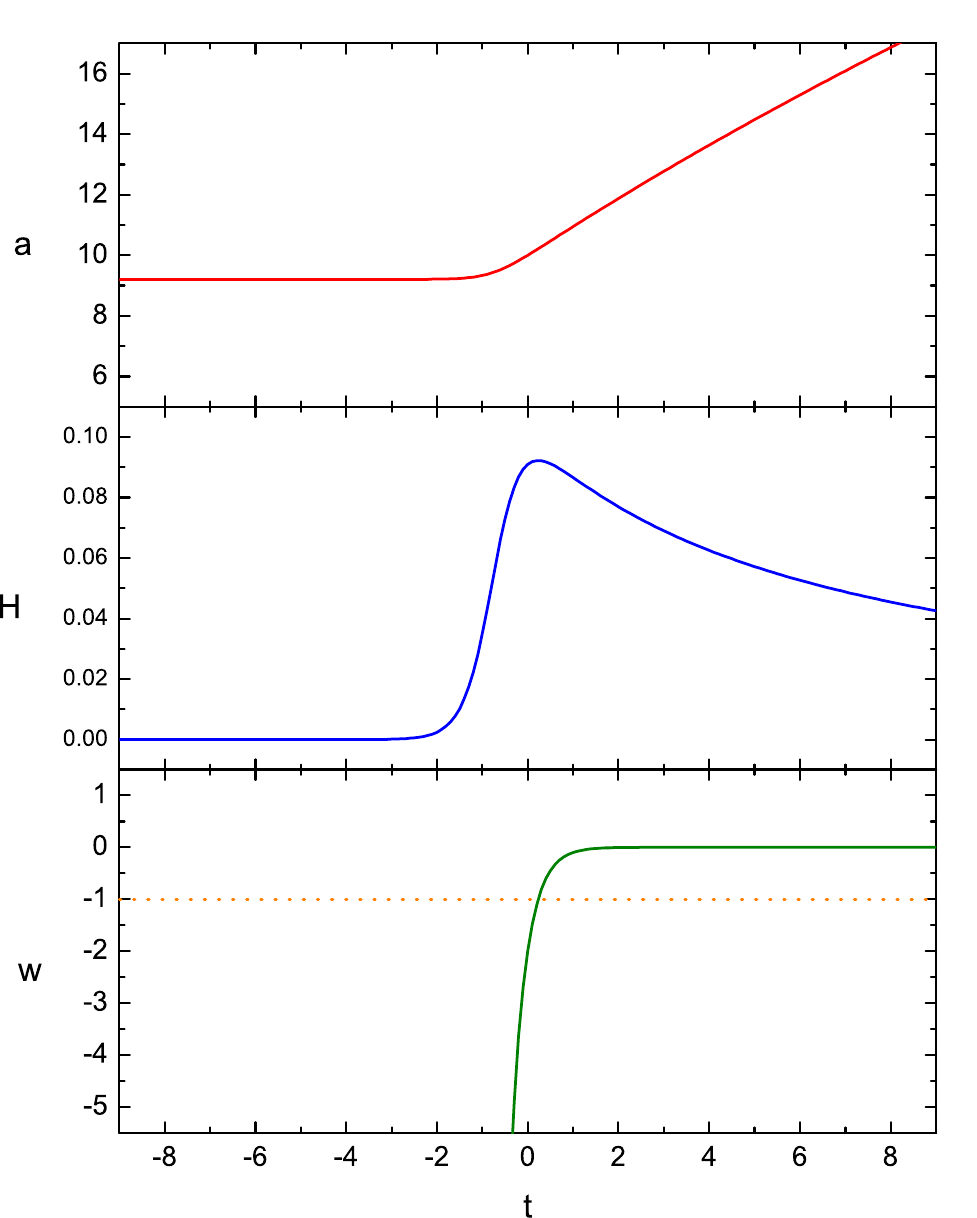}
\caption{Plot of the evolutions of the scale factor $a$, the Hubble parameter $H$ and the background EoS $w$ as functions of cosmic time in the case of spinor Quintom. In the numerical calculation, we take $\alpha=3$ and $C=10$ which are the same as as in the case of Quintom fluid (Fig. \ref{Fig:emergent1}), and $N=100$ for the scalar bilinear $\bar\psi\psi$. All dimensional parameters are of Planck units. }
\label{Fig:spinor1}
\end{figure}

By numerically solving the Friedmann equations, one can solve out the evolution of the scale factor along  cosmic time. We show the evolutions of the scale factor $a$, the Hubble parameter $H$ and the EoS $w$ in Fig. \ref{Fig:spinor1}.

\begin{figure}
\includegraphics[scale=0.4]{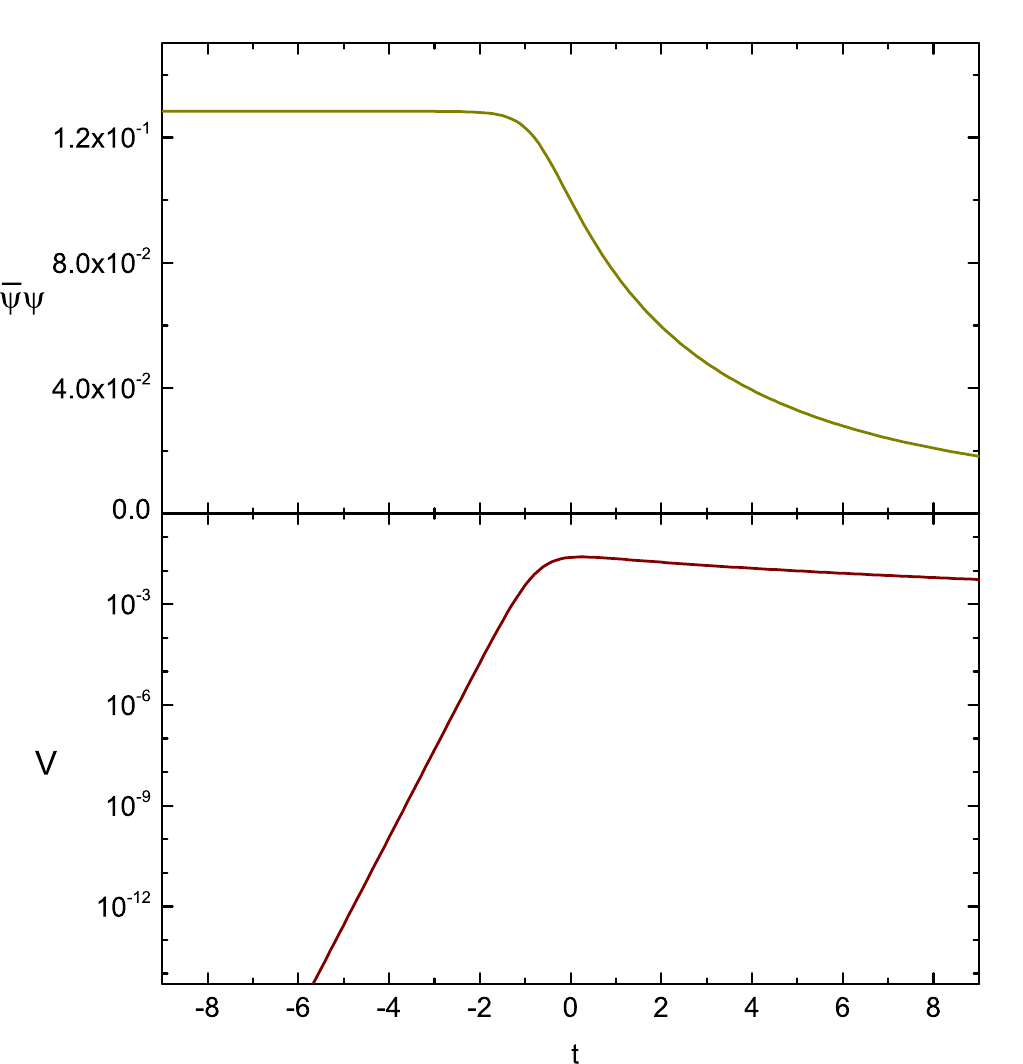}
\caption{Plot of the evolutions of the scalar bilinear $\bar\psi\psi$ and the potential $V$ as functions of cosmic time in the case of spinor Quintom. In the numerical calculation, we take the values of $\alpha$, $C$ and $N$ the same as in Fig. \ref{Fig:spinor1}. All dimensional parameters are of Planck units. }
\label{Fig:spinor2}
\end{figure}

After having known the evolution of the scale factor, one can numerically obtain that of the scalar bilinear $\bar\psi\psi$ due to the relation that $\bar\psi\psi\propto a^{-3}$. In addition, the energy density of the spinor field only depends on the potential $V$ and thus one can derive $V$ evolving along cosmic time. They are shown in Fig. \ref{Fig:spinor2}. From the upper panel of this figure, one can see that $\bar\psi\psi$ is a monotonically decreasing function and approaches a constant in the emergent universe period. Note that, we use log scale to show the wide range of scales for $V$ in the longitudinal coordinate of the lower panel of Fig. \ref{Fig:spinor2}, where it can be seen that $V$ evolves as an exponential form in the phase of quasi-Minkowski.

\begin{figure}
\includegraphics[scale=0.4]{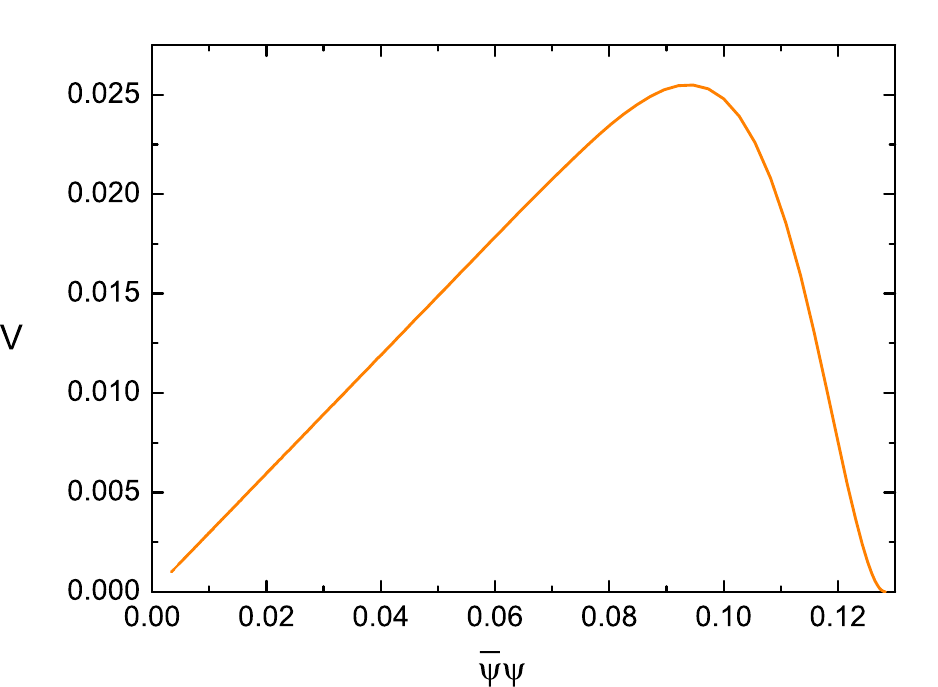}
\caption{Plot of the potential $V$ as a function of $\bar\psi\psi$ in the case of spinor Quintom. In the numerical calculation, we take the values of $\alpha$, $C$ and $N$ the same as in Fig. \ref{Fig:spinor1}. All dimensional parameters are of Planck units. }
\label{Fig:spinor3}
\end{figure}

Since the evolutions of $\bar\psi\psi$ and $V$ were already obtained in above numerics, we can further derive $V$ as a function of $\bar\psi\psi$. This is achievable as $\bar\psi\psi$ is decreasing monotonically along the cosmic time and then can lead to a inverse function $t=t(\bar\psi\psi)$. To combine $t=t(\bar\psi\psi)$ and $V=V(t)$, we then numerically solve out $V=V(\bar\psi\psi)$ as shown in Fig. \ref{Fig:spinor3}.

Interestingly, from Fig. \ref{Fig:spinor3} one can find that $V$ is a linear function of $\bar\psi\psi$ at the regime of small $\bar\psi\psi$ values. This implies that the spinor Quintom recovers the conventional form of a massive fermion with its potential $V \sim m\bar\psi\psi$ at low energy limit. Moreover, in the UV limit, one can derive the asymptotical form of the scale factor as follows,
\begin{eqnarray}\label{a_Emergent}
 a(t) \simeq a_E (1 + Ce^{\alpha M_p t})^{\frac{1}{\alpha C}} ~,
\end{eqnarray}
and thus the asymptotical forms of $\bar\psi\psi$ and $V$. To take the inverse function of $\bar\psi\psi(t)$, eventually we can get the approximate expression of the potential in the phase of quasi-Minkowski:
\begin{eqnarray}\label{V_UV}
 V(\bar\psi\psi) \simeq \frac{3M_p^4}{C^2} \bigg[ 1-a_E^{\alpha C} (\frac{\bar\psi\psi}{N})^{\frac{\alpha C}{3}} \bigg]^2 ~,
\end{eqnarray}
where $a_E$ being the minimal value of the scale factor in the emergent universe scenario.

{\tc{Note that, the potential for the spinor Quintom in low energy limit is very normal in quantum field theory, which is merely a mass term. However, the potential becomes very nontrivial at the high energy scale, which takes the form \eqref{V_UV}. Although this form is purely phenomenologically constructed, one can find that it implies a condensate of the tachyonic spinor field in UV regime and thus is expected to be related to the spinor formalism in open string field theory\cite{Berkovits:1995ab, Berkovits:2001nr}.}}

\section{Conclusion}

As well known, the conception of Quintom scenario was originally from the study of dark energy physics, which shows the EoS of dark energy might cross the cosmological constant boundary \cite{Feng:2004ad}. This scenario is mildly favored by a group of cosmological observations \cite{Xia:2007km, Zhao:2012aw}, and its dynamics were extensively analyzed in a number of works (for example see \cite{Feng:2004ff, Guo:2004fq, Vikman:2004dc, Nojiri:2005sx, Caldwell:2005ai, Li:2005fm, Cai:2006dm, Cai:2007gs} for phenomenological study and see \cite{Cai:2009zp} for a comprehensive review). However, it was soon found that the Quintom behavior has many significant implications to early universe physics. Namely, it can give rise to nonsingular bouncing and cyclic \cite{Xiong:2008ic, Cai:2009in, Cai:2011bs} solutions when applied to high energy scale.

To conclude, we in the current paper have studied the realization of emergent universe scenarios in the presence of Quintom matters. In the literature there have been a lot of efforts in building the models of emergent universe to avoid the big bang singularity. Also there are efforts on investigating the perturbation analysis in the phase of emergent universe assuming the gravity sector is still described by standard Einstein gravity. However, we in the present paper point out that a model giving rise to the emergent universe scenario has to be of Quintom behavior within the standard 4-dimensional FRW framework under the assumption of Einstein gravity.

In explicit realizations, we first considered the Quintom fluid with a parameterized form of EoS to illustrate its possibility. After that, we have studied in detail the model of spinor Quintom. As the model of spinor Quintom possesses a very nice algebra relation in the curved spacetime, we are able to reconstruct a suitably chosen form of the potential and force it to give rise to the emergent universe scenario. Moreover, this type of model is able to exit the phase of quasi-Minkowski expansion gracefully and enter a normal expanding phase dominated by dust matter.

{\tc{The scenario proposed in the current letter and its implementation in the detailed model has some remarkable properties in phenomenological applications.}}

{\tc{First of all, it is interesting to realize the Quintom scenario in other effective field models, such as the Galileon-type fields. In the original model of {\it Galilean Genesis}, it was found that the universe could hardly exit the emergent state smoothly since its equation of state was unable to cross $-1$ from below to above. According to our analysis, one can expect a much improved model of {\it Galilean Genesis} which can take the advantage of Quintom scenario to exit the primordial era smoothly.}}

Additionally, a crucial issue in the cosmology of emergent universe is the processing of cosmic perturbations throughout the quasi-Minkowski expanding phase. In the literature, there have been extensive studies on the generation of primordial power spectrum. As a first step, we in the present paper only considered the background evolution but requires the energy scale of the emergent universe to be much lower than the Planck scale, so that the assumption of Einstein gravity is trustable in our model. However, we should be aware of that since a cosmic spinor was introduced to realize the scenario of emergent universe, it may seed some unwanted fluctuations modes such as vector modes through gauge interactions. Such a complete perturbation analysis of the emergent universe Quintom model lies beyond the scope of the present work and it is left for future investigation.

\begin{acknowledgments}
We thank Robert Brandenberger and Yun-Song Piao for useful discussions. The work of CYF is supported in part by Department of Physics in McGill University and the Cosmology Initiative in Arizona State University. The research of ML is supported in part by National Science Foundation of China under Grants No. 11075074 and No. 11065004, by the Specialized Research Fund for the Doctoral Program of Higher Education (SRFDP) under Grant No. 20090091120054 and by SRF for ROCS, SEM. XZ is supported in part by the National Science Foundation of China under Grants No. 10975142 and 11033005, and by the Chinese Academy of Sciences under Grant No. KJCX3-SYW-N2. One of the author (CYF) is grateful to the other two (ML and XZ) and Yun-Song Piao for hospitality during his visit in Nanjing University and the Institute of High Energy Physics at CAS and the Graduate University of CAS while this work was initiated.
\end{acknowledgments}

\section*{Appendix}

In the first part of this Appendix, we derive the explicit expressions for the energy density and pressure provided in Eqs. \eqref{density} and \eqref{pressure}, respectively. In the second part of the Appendix, we address on the stability of our model throughout the cosmological evolution.

\subsection{Energy density and pressure}

We assume the spinor field is space-independent but evolves along cosmic time. Then the equations of motion given in \eqref{EoMpsi} and \eqref{EoMbarpsi} are simplified as follows,
\begin{eqnarray}
\label{EoM_psi}
 i\gamma^{0} (\dot{\psi} + \frac{3}{2}H\psi) - V' \psi &=& 0~,\\
\label{EoM_barpsi}
 i (\dot{\bar\psi} + \frac{3}{2}H\bar\psi)\gamma^{0} + V' \bar\psi &=& 0~.
\end{eqnarray}

According to the expression of the energy-momentum tensor given in Eq. (\ref{EMT}), we calculate the expression of the energy density of the spinor field:
\begin{eqnarray}
 \rho_{\psi} &=& T_{0}^{0} \nonumber\\
 &=& \frac{i}{4} [\bar\psi\Gamma^{0}D_{0}\psi +\bar\psi \Gamma_{0}D^{0}\psi -D_{0}\bar\psi\Gamma^{0}\psi -D^{0}\bar\psi\Gamma_{0}\psi] \nonumber\\
 &&~ - [\frac{i}{2}(\bar\psi\Gamma^{\mu}D_{\mu} \psi-D_{\mu}\bar\psi\Gamma^{\mu}\psi)-V(\bar\psi\psi)] \nonumber\\
 &=& \frac{i}{2a}(\bar\psi\gamma^{i}\partial_{i}\psi -\partial_{i}\bar\psi\gamma^{i}\psi) +V(\bar\psi\psi) \nonumber\\
 &=& V(\bar\psi\psi)~,
\end{eqnarray}
where in the last equality we used the assumption of scale independence $\partial_i\psi=0$. Furthermore, we can compute the pressure of the spinor field:
\begin{eqnarray}
 p_{\psi}&=&-T_{i}^{i} \nonumber\\
 &=& \frac{i}{2}(\bar\psi\gamma^{0}D_{0}\psi-D_{0}\bar\psi\gamma^{0}\psi) -\frac{i}{2a}(\bar\psi\gamma^{i}\partial_{i}\psi -\partial_{i}\bar\psi\gamma^{i}\psi) \nonumber\\
 &&~ -V(\bar\psi\psi) \nonumber\\
 &=& \frac{i}{2}(\bar\psi\gamma^{0}D_{0}\psi-D_{0}\bar\psi\gamma^{0}\psi) -V(\bar\psi\psi) \nonumber\\
 &=& V'\bar\psi\psi-V ~,
\end{eqnarray}
where we have used the assumption $\partial_i\psi=0$ in the second line and applied Eqs. \eqref{EoM_psi} and \eqref{EoM_barpsi} in the last line.

\subsection{The stability issue}

Regarding the stability issue of a cosmological model, there are mainly two possible concerns. One is to check whether or not the model leads to a ghost degree of freedom. Obviously, this issue does not exist in our emergent universe model which is realized by a spinor Quintom. As one can see from the action, the kinetic term for the spinor field is very standard without any modification. Thus, there does not exist any extra degree of freedom in our model. The other is to study the evolution of cosmological perturbations seeded by the spinor field and make sure the backreaction of perturbations to the background is controllable.

Here we would like to show the perturbation theory of Spinor Quintom crudely. In order to simplify the derivative, we would like to redefine the spinor as $\psi_N\equiv a^{\frac{3}{2}}\psi$, and perturb it as
\begin{eqnarray}
 \psi_N\rightarrow\psi_N(t)+\delta\psi_N(t, x^i)~,
\end{eqnarray}
around the homogeneous background. Then perturbing the equation of motion of the spinor field, one can obtain the perturbation equation as follows,
\begin{eqnarray}\label{perteq}
&\frac{d^2}{d\tau^2}\delta\psi_N -\nabla^2\delta\psi_N +a^2 [V'^2+i\gamma^0 (HV'-3HV''\bar\psi\psi)] \delta\psi_N
 \nonumber\\
& = -2a^2V'V''\delta(\bar\psi\psi)\psi_N -i\gamma^\mu\partial_\mu[a
V''\delta(\bar\psi\psi)]\psi_N~,
\end{eqnarray}
where $\tau$ is the conformal time defined by $d\tau\equiv dt/a$. Since the right hand side of the equation decays proportional to $a^{-3}$ or even faster, we can neglect those terms throughout the evolution of the universe for simplicity.

From the perturbation equation above, we can read that the sound speed is equal to $1$ which eliminates the instability of the
system in short wavelength. Moreover, when the equation of state $w$ crosses $-1$, we have $V'=0$ at that moment and the eigen function of the solution to Eq. (\ref{perteq}) in momentum space is a Hankel function with an index $\frac{1}{2}$. Therefore, the perturbations of the spinor field oscillate inside the hubble radius. This is an interesting result, because in this way we might be able to establish the quantum theory of the spinor perturbations, just as what is done in inflation theory.

Moreover, in the emergent universe phase one can insert the expressions of scale factor \eqref{a_Emergent} and Hubble parameter \eqref{H_Emergent}, and apply the approximate form of the potential \eqref{V_UV} into the perturbation equation \eqref{perteq}. Then one can get $H\sim e^{\alpha M_pt}$ and $V'\sim e^{\alpha M_pt}$ which are exponentially suppressed but $V''\sim {\rm constant}$. This implies that the perturbations of the spinor field are almost vacuum fluctuations in the phase of emergent universe which take the form of plane wave function approximately. As a consequence, the backreaction of perturbations are negligible in our model.

\end{document}